\documentclass[10pt,conference]{IEEEtran}

\usepackage{cite}
\usepackage{pslatex}
\usepackage{epsfig}
\usepackage{amsmath}
\usepackage{amssymb}
\usepackage{array}
\usepackage{multirow}
\usepackage{threeparttable}
\usepackage{url}
\usepackage{dsfont}
\usepackage{color}

\newtheorem{definition}{Definition}

\renewcommand{\QED}{\QEDopen}
\renewcommand{\thefootnote}{\fnsymbol{footnote}}

\DeclareMathAlphabet{\mathpzc}{OT1}{pzc}{m}{it}
\renewcommand{\baselinestretch}{1}
\IEEEoverridecommandlockouts 

\def\be{{\mathbf e}}

\def\bv{{\mathbf v}}
\def\bw{{\mathbf w}}
\def\bx{{\mathbf x}}
\def\by{{\mathbf y}}

\def\bE{{\mathbf E}}

\def\b0{{\mathbf 0}}

\begin{document}

\renewcommand{\baselinestretch}{0.95}
\IEEEoverridecommandlockouts 

\title{Graph-based Code Design for Quadratic-Gaussian Wyner-Ziv Problem with Arbitrary Side Information}

\author{
\authorblockN{Yi-Peng Wei*, Shih-Chun Lin\dag, Yu-Hsiu Lin* and Hsuan-Jung Su*}
\authorblockA{ *Graduate Institute of Communication Engineering,
National Taiwan University, Taipei, Taiwan  \\
\dag Graduate Institute of Computer and Communication
Engineering, National Taipei University of Technology, Taipei, Taiwan \\
r98943073@ntu.edu.tw, sclin@ntut.edu.tw, r97942045@ntu.edu.tw,
hjsu@cc.ee.ntu.edu.tw}}

\maketitle 

\renewcommand{\thefootnote}{\fnsymbol{footnote}}

\begin{abstract}
Wyner-Ziv coding (WZC) is a compression technique using
decoder side information, which is unknown at the encoder, to help the
reconstruction. In this paper, we propose and implement a new WZC
structure, called \textit{residual WZC}, for the quadratic-Gaussian
Wyner-Ziv problem where side information can be arbitrarily
distributed. In our two-stage residual WZC, the source is
quantized twice and the input of the second stage is the
quantization error (residue) of the first stage. The codebook of
the first stage quantizer must be simultaneously good for source
and channel coding, since it also acts as a channel code at the
decoder. Stemming from the non-ideal quantization at the encoder, a problem
of channel decoding beyond capacity is identified and solved when
we design the practical decoder. Moreover, by using the
modified reinforced belief-propagation quantization algorithm,
the low-density parity check code (LDPC), whose edge degree is optimized
for channel coding, also performs well as a source code. We then implement the residual WZC by an LDPC and a low-density generator matrix code (LDGM).
The simulation results show that our practical construction approaches the Wyner-Ziv bound. Compared with previous works, our construction can offer more design flexibility in terms of distribution of side information and practical code rate selection.
\footnote[0]{This work was supported by National Science Council,
National Taiwan University and Intel Corporation under Grants
NSC 100-2911-I-002-001, NSC 100-2218-E-027-008 and 10R70501.}
\end{abstract}

\section{Introduction}
Wyner-Ziv coding (WZC) \cite{Wyner-t2}, which is a generalization
of the lossless Slepian-Wolf coding (SWC) \cite{Slepian-n}, refers to
a lossy source coding with side information at the decoder only.
Among the applications of WZC \cite{Xiong_SPM_04}, one important scenario
is the quadratic-Gaussian problem, where the difference between the source
and the side information is Gaussian and mean-square error (MSE) is adopted
as the distortion measure.
For this case, WZC does not incur rate loss when compared with
source coding with side information available at both the encoder
and the decoder. This zero-rate-loss result was originally proved
for the case where the side information is Gaussian \cite{Wyner-t2},
but recently generalized to the case where the side information can
be arbitrarily distributed \cite{Zamir-n}.
There are many applications of quadratic-Gaussian WZC, such as distributed source
coding in the sensor networks \cite{Xiong_SPM_04} and the protocols in the
relay networks \cite{Kramer-c}.

The theoretical analysis of Wyner and Ziv \cite{Wyner-t2} was based
on the non-structured codebook which is hard to implement in practice.
Later on, some theoretical works based on structured codebooks were proposed
\cite{Zamir-n}\cite{SCS_IT_09}\cite{Korada_IT_10}. However, only
\cite{Zamir-n} focused on the quadratic-Gaussian cases. Whether the theoretical results in
\cite{SCS_IT_09}\cite{Korada_IT_10} can be applied to such cases
is unknown. Moreover, the results in
\cite{Zamir-n}\cite{SCS_IT_09}\cite{Korada_IT_10} require two
nested codebooks good for the source and/or channel coding.
Practically constructing such good codebooks with nested
structure in \cite{Zamir-n}\cite{SCS_IT_09}\cite{Korada_IT_10}
is still a challenging task.

Instead of building WZC from nested codebooks as in
\cite{Zamir-n}\cite{SCS_IT_09}\cite{Korada_IT_10}, we proposed a
new coding structure in our previous work \cite{Lin_ITW_09}, where
two codebooks \textit{without} nested structure were used in a
two-stage serial quantization process. The encoder first quantizes
the source once, and then uses another codebook to quantize the
quantization error of the first stage. The quantization index of
the second stage is then sent to the decoder. We name the coding
in \cite{Lin_ITW_09} residual WZC, which reflects some
resemblances of our coding structure to the residual vector
quantizer \cite[Sec 12.11]{gersho1992vector} for source coding
without side information. In a theoretical random coding setting,
the residual WZC was proved to be Wyner-Ziv-bound achieving.

In this paper, we show that using graph-based codes, the
theoretically-optimal residual WZC in \cite{Lin_ITW_09} can be
practically implemented. Different from the celebrated
quadratic-Gaussian WZC in \cite{Xiong_TCOM_09}\cite{Xiong_IT_06},
our scheme can approach the Wyner-Ziv bound for \textit{all rate
regimes when the side information is arbitrarily distributed}. By using
the loss to the Wyner-Ziv bound as the performance metric, simulation shows that
the performance of our code is comparable to that in \cite{Xiong_TCOM_09}.
However, in contrast to \cite{Xiong_TCOM_09}, our simulation result is
independent of the distribution of the side information, as long as the
variance of the side information is the same.
Besides, our WZC has a complexity similar to that of \cite{Xiong_TCOM_09} and
linear in the codeword length, and much lower than that of the
lattice decoder (NP problem) in \cite{Zamir-n}.

Practically implementing the residual WZC is not trivial.
Firstly, the codebook of the first stage quantizer acts
as a channel code in our decoder, and must be simultaneously good
for source and channel coding (SSC). Moreover, the non-ideal
practical quantizers at the encoder make the channel decoder
operate in a rate regime above capacity. We propose a method
which can increase the equivalent signal-to-noise ratio (SNR) at
the channel decoder to solve this problem.
Secondly, although the low-density parity check code (LDPC) has been
proved to be SSC theoretically \cite{Chandar_thesis_10}, the edge degree design and
practical quantization algorithm for SSC LDPC are still unknown.
We modify the recently proposed reinforced belief propagation
(RBP) algorithm \cite{Braunstein_arXiv_11} so that the LDPC,
whose edge degree is originally optimized for channel coding, is good
for source coding. The codebook of the second stage of our
residual WZC is chosen as a low-density generator matrix code
(LDGM). We also apply the RBP algorithm to LDGM, which is much
simpler than the well-known hard-decimation-based algorithm
\cite{Filler_LDGM} for LDGM, and reduces the overall WZC complexity
a lot. More detailed comparisons between our practical
construction and previous works can be found in Section
\ref{sec_discuss}.

\section{System model and residual Wyner-Ziv coding} \label{secWZC}
In the considered quadratic-Gaussian Wyner-Ziv problem, the
relationship between the length-$n$ source vector $\mathbf{x}$ and the arbitrary
side information $\mathbf{y}_a$ is \footnote{In this paper, a
vector is denoted in bold lower-case, while the superscript
$T$ denotes the transpose of a vector. A zero-mean Gaussian random variable
with variance $\sigma^2$ is denoted by $N(0,\sigma^2)$. A random variable $X$ for Shannon's random coding setting
is denoted in capital italic.  The entropy is denoted as $h(\cdot)$. All logarithms are of base 2.}
\vspace{-0.1cm}
\begin{equation}\label{system_model}
\mathbf{x}=\mathbf{y}_a+\mathbf{v},
\end{equation}
where $\mathbf{v}$ is an independent and identically distributed (i.i.d.)
Gaussian vector with each element distributed as $N(0,P_V)$, and $\mathbf{v}$ is
independent of $\mathbf{y}_a$. Moreover, $\mathbf{y}_a$ can be
arbitrarily distributed. Denoting the reconstruction as
$\mathbf{\hat{x}}$, the ``quadratic'' MSE distortion measure is
adopted with maximum distortion $D$
\vspace{-0.1cm}
\begin{equation} \label{distortion_measure}
\frac{1}{n}\bE[(\mathbf{x}-\mathbf{\hat{x}})^2]\leq D.
\end{equation}
The Wyner-Ziv bound is then \cite{Zamir-n}
\begin{equation} \label{eq_WZ_rate}
R_{WZ}(D)=\frac{1}{2}\log\left(\frac{P_V}{D}\right),
\end{equation}
when $P_V>D$. When $P_V \leq D$, $R_{WZ}(D)=0$ and we will neglect
this trivial case in the following.
\vspace{-0.2cm}



\begin{figure}
\centering \epsfig{file=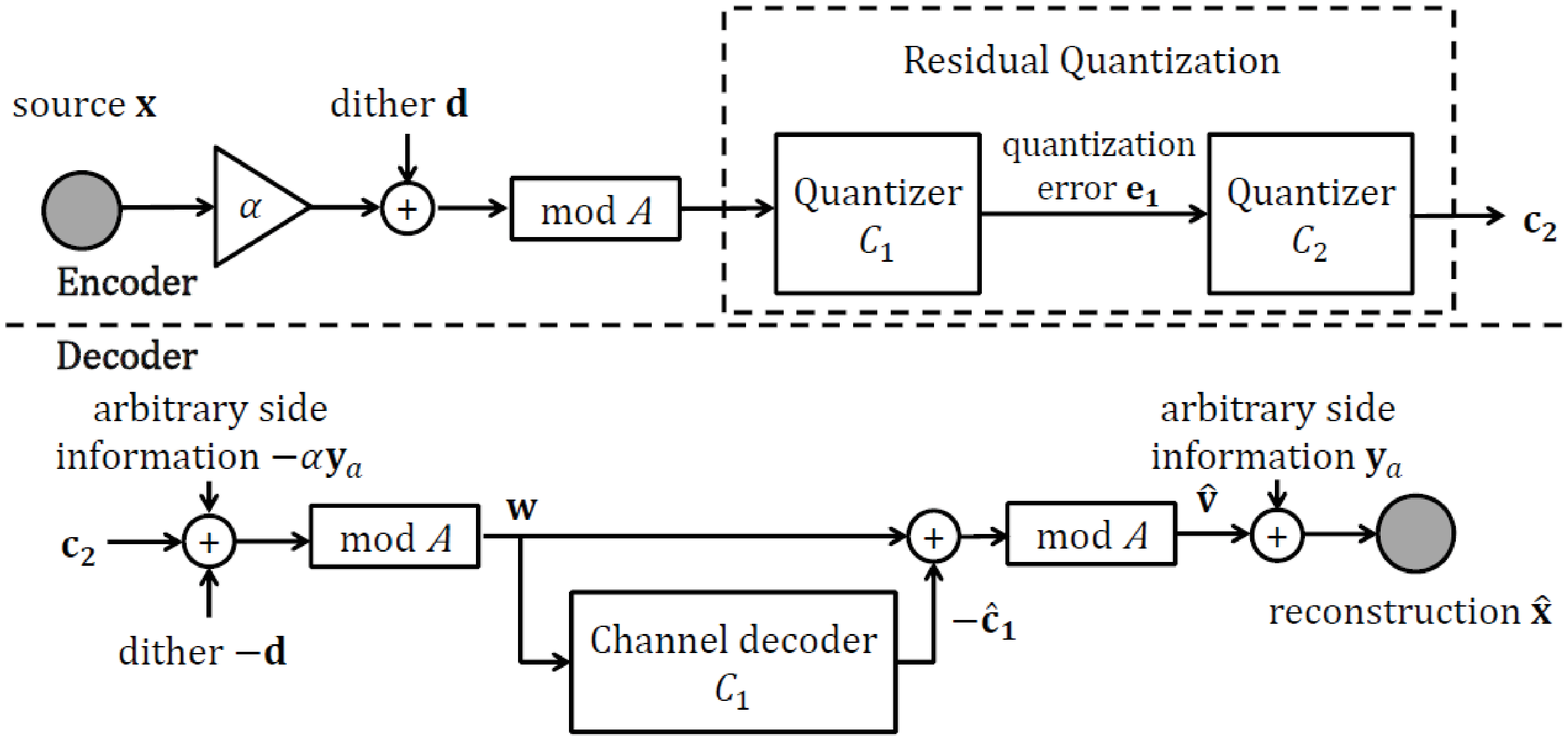
,width=0.5\textwidth} \caption{Coding structure of the proposed residual
Wyner-Ziv coding} \label{SQ_structure}
\vspace{-0.55cm}
\end{figure}

\subsection{Structure of the proposed residual Wyner-Ziv coding}
We briefly review the residual WZC in \cite{Lin_ITW_09}, and the coding structure is depicted in Fig.~\ref{SQ_structure}.
First, we introduce some definitions as follows

\begin{definition}
Given a vector
$\mathbf{x}=(x_{1},\ldots,x_{n})^{T}$ and a predetermined value $A$,
the modulo $A$ operation is defined as
\[
{\mathbf{x} \bmod A}  
={(x_{1}-Q_{A}(x_{1}),\ldots,x_{n}-Q_{A}(x_{n}))^{T}},
\]
where $Q_{A}(x_{i})$ is the nearest multiple of $A$ to $x_{i}$,
$\forall i$.
\end{definition}

\begin{definition}[mod-A distance] \label{Def_modA_dis}
The mod-$A$ distance between two vectors
$\mathbf{a}=(a_{1},\ldots,a_{n})^{T}$ and
$\mathbf{b}=(b_{1},\ldots,b_{n})^{T}$ is
\vspace{-0.1cm}
 \[
{||{\mathbf{b}}-{\mathbf{a}}||^{2}_{A} \equiv \sum_{i=1}^n
[(b_i-a_i) \bmod A]^{2}}.
\]
\end{definition}

The residual WZC is described in detail as follows:

\noindent \textbf{Encoder part:} The input of the first-stage quantizer $C_{1}$ in Fig.~\ref{SQ_structure} is
\[
(\alpha \mathbf{x}+\mathbf{d}) \bmod A,
\]
where $\alpha$ is a scaling factor which
will be determined later, $\mathbf{d}$ is a randomly generated dither
signal known both at the encoder and the decoder, and $A$ is called as the modulo size.
The entries of the dither are uniformly distributed in the interval
$[-{A \over 2} ,{A \over 2}]$. According to Crypto Lemma in \cite{Lin_ITW_09}\cite{Zamir-n}, we know
that ($\alpha \mathbf{x}+\mathbf{d}$) mod $A$ is uniform in the
region $[-{A \over 2} ,{A \over 2}]^n$ and independent of $\bx$. Then the distribution of $\bx$, determined by the distribution of side information $\by_a$ from \eqref{system_model}, will not affect the quantization result of our residual WZC encoder. The first-stage quantizer $C_1$ searches a codeword $-\mathbf{c_1}$ in $C_1$
such that the mod-$A$ distance (in Defintion \ref{Def_modA_dis}) between $\alpha\mathbf{x}+\mathbf{d}$ and
$-\mathbf{c_1}$ is minimized, under the distortion constraint
$
\frac{1}{n}\bE[\mathbf{e_1}^T\mathbf{e_1}]\leq\alpha D+\alpha ^2 P_V.
$
The quantization error after the first stage is
\begin{equation} \label{eq_error_1}
\mathbf{e_1}=(\alpha\mathbf{x}+\mathbf{d}+\mathbf{c_1})\bmod A.
\end{equation}

The input of the second-stage quantizer $C_2$ in Fig.~\ref{SQ_structure} is the quantization error $\mathbf{e_1}$ of the first stage in \eqref{eq_error_1}. The distortion constraint for the second stage is
$
\frac{1}{n}\bE[\mathbf{e_2}^T\mathbf{e_2}]\leq\alpha D,
$
and the quantization output and the quantization error of the second stage are  $\mathbf{c_2}$ and
\vspace{-0.2cm}
\begin{equation} \label{eq_error_2}
\mathbf{e_2}=(\mathbf{e_1}-\mathbf{c_2})\bmod A,
\end{equation}
respectively. Finally, the encoder sends the index representing
$\mathbf{c_2}$ to the decoder.

\noindent \textbf{Decoder part:} The decoder receives the index
representing $\mathbf{c_2}$ and the side information $\mathbf{y}_a$.
As in Fig.~\ref{SQ_structure}, the decoder first computes
${\mathbf{w}} = {(\mathbf{c_2}- \alpha \mathbf{y}_a
-\mathbf{d})\bmod A}$. From \cite{Lin_ITW_09}, equivalently we
have
\begin{equation} \label{equivalent_channel}
\bw= {(\mathbf{c_1}+\alpha \mathbf{v} - \mathbf{e_2}) \bmod A}.
\end{equation}
Then we can channel decode $\mathbf{c_1}$ from $\bw$, by treating
\vspace{-0.1cm}
\begin{equation} \label{eq_equivalent_noise}
\alpha\mathbf{v}-\mathbf{e_2}
\end{equation}
\vspace{-0.1cm}
as the equivalent channel noise, where $\mathbf{e_2}$ is given
in \eqref{eq_error_2}. By denoting the channel decoder output as
$\mathbf{\hat{c}_1}$, we can compute $\mathbf{\hat{v}}$, the
reconstruction of $\bv$, as $ {\mathbf{\hat{v}}} =
{(\mathbf{w}-\mathbf{\hat{c}_1})\bmod A }\nonumber $. Finally, the
reconstruction $\mathbf{\hat{x}}$ is
\[
\mathbf{\hat{x}}=\mathbf{y}_a+\mathbf{\hat{v}}.
\]

Let the code rates of $C_1$ and $C_2$ be $R_1$ and $R_2$,
respectively. According to \cite{Lin_ITW_09}, by letting $\alpha$ be
\begin{equation}\label{alpha_formula}
\alpha=1-D/P_V,
\end{equation}
there exists codebooks $C_1$ and $C_2$ such that the MSE
distortion constraint (\ref{distortion_measure}) is met if
\begin{align}
R_1  \geq  \log A-\frac{1}{2}&\log (2\pi e (\alpha P_V))+\epsilon
_1, R_2  \geq \frac{1}{2}\log (\frac{P_V}{D})+\epsilon _2
\label{Rq_enc} \\
R_1 & \leq \log A-\frac{1}{2}\log (2\pi e (\alpha P_V))+\epsilon
_1 \label{Rq_dec},
\end{align}
where $\epsilon_1,\epsilon_2 \rightarrow 0$ as $A \rightarrow
\infty$ and the code length $n \rightarrow \infty$. Indeed
\eqref{Rq_enc} ensures the success of the two quantization processes
in the encoder, while \eqref{Rq_dec} ensures the success of
channel decoding in the decoder \cite{Lin_ITW_09}. Then from
\eqref{Rq_enc}\eqref{Rq_dec}, we know that one can achieve the
Wyner-Ziv bound in \eqref{eq_WZ_rate} as
\begin{equation} \label{R_WZC}
R_2=\frac{1}{2} \log (\frac{P_V}{D}).
\end{equation}

\section{Design flow for graph-based code implementation}
\begin{figure}
\centering \epsfig{file=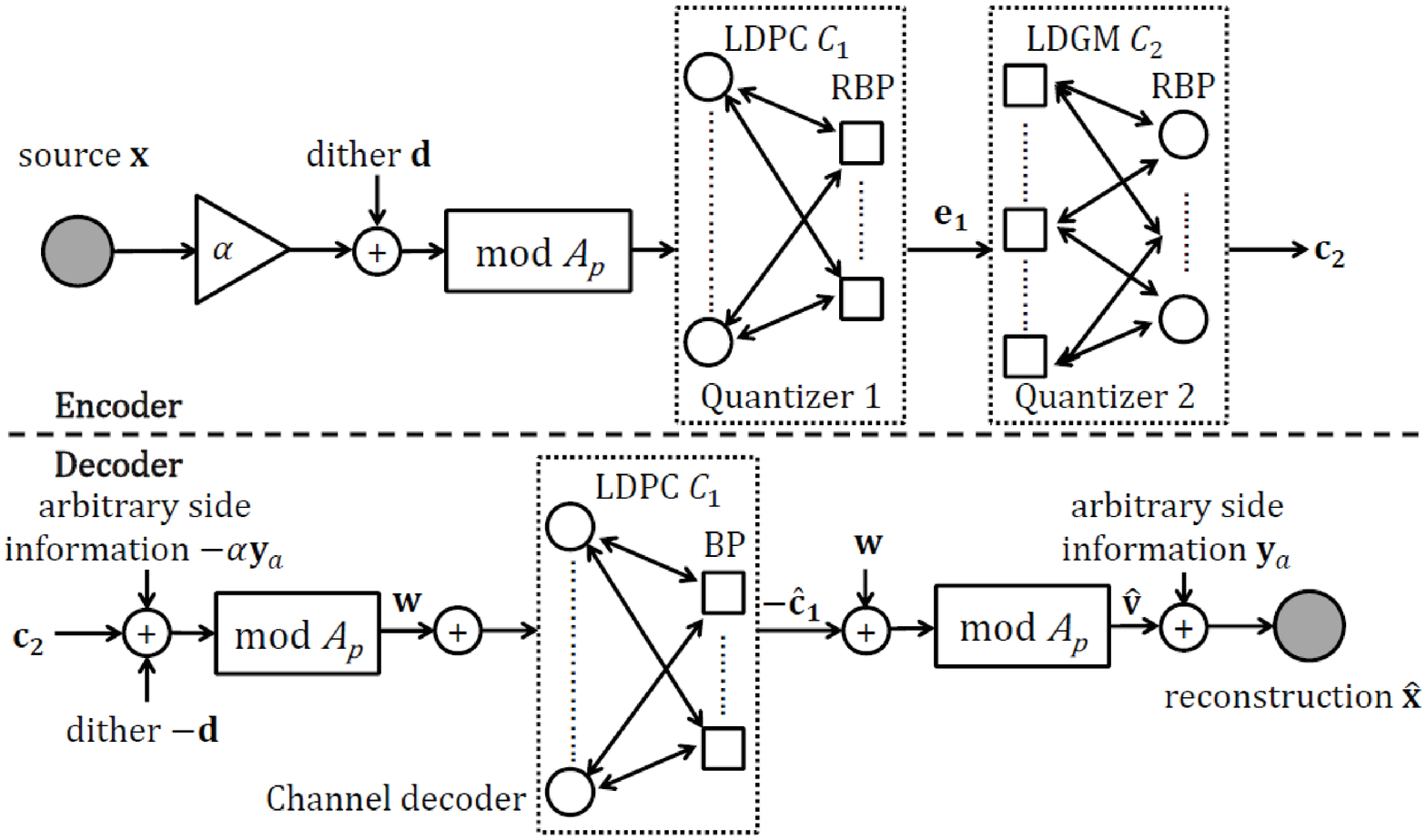
,width=0.5\textwidth} \caption{Graph-based code implementation of
residual Wyner-Ziv coding} \label{graph}
\vspace{-0.55cm}
\end{figure}

Theoretically, from \cite{Lin_ITW_09}, the $C_1$ must be a SSC
code while $C_2$ must be a good source code. We implement $C_1$
and $C_2$ in Fig.~\ref{SQ_structure} by an LDPC and an LDGM,
respectively. The detailed block diagram of our practical
implementation is shown in Fig.~\ref{graph}. Before introducing
our flow to select the code parameters and the detailed
encoding/decoding algorithms in Section \ref{sec_flow}, in
Section \ref{Sec_problems}, we first discuss some problems
encountered when building our flow for the practical implementation.
\vspace{-0.2cm}
\subsection{Issues for practical implementation} \label{Sec_problems}
\noindent \textbf{Finite modulo size : } In \cite{Lin_ITW_09}, the
Wyner-Ziv bound is achieved while the modulo size $A$ approaches
infinity. However, infinite modulo size $A$ is impractical since the
rate $R_1$ in \eqref{Rq_enc} will become infinite, too. Thus we propose the following method to estimate how large $A$ should be in practice.
From \cite{Lin_ITW_09}, the $\epsilon_1$ and $\epsilon_2$ in \eqref{Rq_enc} satisfy
\begin{equation} \label{eq_pra_A1}
\epsilon_2<\epsilon_1 = h(\alpha V-E_2)-h((\alpha V-E_2) \bmod A),
\end{equation}
where $V \sim N(0,P_V)$, and $E_2$ is a random variable corresponding to $\mathbf{e_2}$ in \eqref{eq_error_2} in the Shannon-random coding setting. The distribution of $E_2$ is detailed in \cite{Lin_ITW_09}. By numerically calculating the right-hand-side (RHS) of the equality in \eqref{eq_pra_A1}, we can choose  $A$ to make $\epsilon_1$ and $\epsilon_2$ smaller than a threshold $\epsilon$.

In the following, we use the modulo size $A_\epsilon$ to
represent the sufficiently large modulo size such that
we can essentially neglect $\epsilon_1$ and $\epsilon_2$ in \eqref{Rq_enc}. \\
\vspace{-0.2cm}

\noindent \textbf{Channel decoding beyond capacity :} With $A_\epsilon$ obtained previously, we can select $R_1$ from \eqref{Rq_enc} by replacing $A$ with $A_\epsilon$ and neglect $\epsilon_1$. However, if we directly implement the channel decoder in Fig.~\ref{graph} with selected $R_1$ and $A_\epsilon$, the channel decoder will always fail. The reason is as follows. Since no practical quantizers $C_1$ and $C_2$ can exactly achieve the rate-distortion bound, we will get the practical quantization error
variance $D_{2,\epsilon}$ of $\be_2$ in \eqref{eq_error_2} larger than the theoretical one predicted in \cite{Lin_ITW_09}. From \eqref{equivalent_channel}, the variance of the equivalent noise \eqref{eq_equivalent_noise} will also be larger than the theoretical value in \cite{Lin_ITW_09}, which makes the selected $R_1$ operating in the regime above the channel capacity in \eqref{Rq_dec} with $A=A_\epsilon$.

To solve the channel decoding beyond capacity problem described
above, we propose a method which increases $A_\epsilon$ to $A_p$
while $R_1$ is fixed as that in Step 2 of Table
\ref{tab:AlgBasic}. The key idea is that the equivalent SNR
at the channel decoder in Fig.~\ref{graph} will also increase.
To be more specific, from the
random codebook construction in \cite{Lin_ITW_09}, we know that
the optimal constellation of $C_1$ is uniformly distributed in
$[-\frac{A_\epsilon}{2}, \frac{A_\epsilon}{2}]$. Thus we use the
uniform pulse amplitude modulation (PAM) as the constellation of
$C_1$, and the equivalent SNR of the channel
\eqref{equivalent_channel} with $A=A_\epsilon$ is
\begin{equation} \label{eq_SNR_eps}
SNR_\epsilon=\frac{A_\epsilon^2/k}{D_{2,\epsilon}+\alpha^2P_V},
\end{equation}
where $k$ is a constant related to the order of the PAM constellation.
If we increase the modulo size from $A_\epsilon$ to $A_p$ with rates $R_1$ and $R_2$ unchanged, the quantization error variance of $\be_2$ in \eqref{eq_error_2} will become
\begin{equation} \label{practical_enc_var}
D_{2,p}=(A_p/A_\epsilon)^2D_{2,\epsilon},
\end{equation}
because the variance of $C_1$'s input, which is uniformly-distributed, will increase from $A_\epsilon^2/12$ to $A_p^2/12$. Then the equivalent SNR becomes
\[
SNR_p = \frac{A_p^2/k}{D_{2,p}+\alpha^2P_V}
      = \frac{A_\epsilon^2/k}{D_{2,\epsilon}+(\frac{A_\epsilon}{A_p})^2\alpha^2P_V},
\]
which is larger than $SNR_\epsilon$ in \eqref{eq_SNR_eps} since $A_p>A_\epsilon$.

Now we consider the loss of the practical channel decoder with
practical $C_1$ in Fig.~\ref{graph} compared to the optimal
channel coding in \cite{Lin_ITW_09}. For the practical channel coding,
let $\sigma_{n,\epsilon}^2$ be the variance of the maximum
tolerable equivalent noise for the successful decoding, where the PAM
constellation of $C_1$ is chosen according to $A_\epsilon$. Then
under the non-ideal channel decoding, we have
\vspace{-0.1cm}
\begin{equation} \label{eq_decoder_loss}
\sigma_{n,\epsilon}^2 < D_{2,\epsilon}+\alpha^2P_V,
\end{equation}
where the RHS is the variance of the equivalent noise
\eqref{eq_equivalent_noise}. When the PAM constellation of $C_1$
is chosen according to  $A_p$ instead of $A_\epsilon$, the signal
power of codeword $\mathbf{c_1}$ is scaled by
$(A_p/A_\epsilon)^2$. Then the maximum tolerable equivalent
noise variance $\sigma_{n,p}^2$ for the practical channel coding can
be estimated by
\begin{equation} \label{practical_dec_var}
\sigma_{n,p}^2=(A_p/A_\epsilon)^2 \sigma_{n,\epsilon}^2.
\end{equation}
Now we wish $\sigma_{n,p}^2 \geq D_{2,p}+\alpha^2P_V$ to ensure the
practical channel decoding in \eqref{equivalent_channel} being
successful. Using this criterion with (\ref{practical_enc_var}) and
(\ref{practical_dec_var}), the practical modulo size $A_p$ in
Fig.~\ref{graph} must meet,
\vspace{-0.1cm}
\begin{equation} \label{Ap}
A_p \geq (\frac{A_\epsilon^2\alpha^2P_V}{\sigma_{n,\epsilon}^2-D_{2,\epsilon}})^{\frac{1}{2}}.
\vspace{-0.1cm}
\end{equation}
$\sigma_{n,\epsilon}^2$ and $D_{2,\epsilon}$ can be obtained via numerical simulations. From \eqref{eq_decoder_loss}, we know that $A_p$ selected according to \eqref{Ap} is lager than $A_\epsilon$. Although increasing the modulo size can solve the channel decoding
beyond capacity problem, there will be a loss in the final distortion compared with the
theoretically predicted $D$ using $R_2$ in \eqref{R_WZC} owing to the increment of the
quantization error variance of $\be_2$ in \eqref{eq_error_2} from
\eqref{practical_enc_var}. \\
\vspace{-0.2cm}

\noindent \textbf{Edge degree for SSC LDPC :} One of the critical
parts of our residual WZC is the requirement of a practical SSC
code $C_1$. The best known SSC so far is the trellis coded
quantation (TCQ)/modulation. However, according to the simulation
results in \cite{Chen-s}, using TCQ as our $C_1$ will result in
significant performance loss compared with the Wyner-Ziv bound.
Alternatively, in \cite{Chandar_thesis_10}, LDPC was proved to be an
SSC code for the binary source/channel under the optimal
encoder/decoder. However, the theoretical proof in
\cite{Chandar_thesis_10} gives no hints for the edge degree design and the
practical quantization algorithms for SSC LDPC.
Recently, the RBP algorithm was proposed in \cite{Braunstein_arXiv_11}
for the quantization of the binary source with LDPC. However, the LDPC
in \cite{Braunstein_arXiv_11} is \textit{not} SSC, since its edge degree
exhibits the ultra-sparse structure which makes LDPC poor for the
channel coding when the belief-propagation (BP) algorithm
\cite{mackay1999good} is used.

To solve the practical design problem for SSC LDPC, we propose
using LDPC with the edge degree optimized for the channel coding, and
modifying the binary RBP in \cite{Braunstein_arXiv_11} for our continuous
LDPC quantization with mod-A operation. Although our edge degree for the
first-stage quantizer $C_1$ in Fig.~\ref{graph} is sub-optimal for the
source coding, our simulation results show that it suffices to make the overall
residual WZC approach the Wyner-Ziv bound.
Indeed, our simulation results in Section
\ref{sec_simulation} show that our LDPC has a shaping loss
\cite{gersho1992vector} about only 0.5 dB away from the
rate-distortion bound when the quantizer input is
uniformly-distributed. Our modified RBP algorithm is given in Sec.
\ref{sec_flow}.
\vspace{-0.2cm}

\subsection{Design flow and encoding/decoding algorithms}
\label{sec_flow}

 Our overall design flow is summarized in Table
\ref{tab:AlgBasic} and is explained in detail as follows. The Step
1 and 2 in Table \ref{tab:AlgBasic} are described previously in
Sec. \ref{Sec_problems}. To design the edge degree of the SSC LDPC
in Step 3, as discussed in Sec. \ref{Sec_problems}, we adopt the
EXIT chart fitting approach in \cite{Brink_TCOM_04} to obtain an
LDPC good for the channel coding. By running the BP channel decoding
algorithm, we can calculate the $\sigma^2_{n,\epsilon}$ described
before \eqref{eq_decoder_loss} for our LDPC with the PAM constellation
points. For Step 4, the rate of the LDGM is given in \eqref{R_WZC},
and the quantization alphabet is uniformly spaced with energy
$\alpha^2P_V$ suggested in \cite{Lin_ITW_09}. The edge degree of
LDGM is obtained similarly as in \cite{wang2007approaching}.

To calculate $D_{2,\epsilon}$ in Step 4 of Table
\ref{tab:AlgBasic}, we modify the RBP in
\cite{Braunstein_arXiv_11} as the quantization algorithm for both
LDPC and LDGM at the WZC encoder in Fig.~\ref{graph}. The RBP
is a generalization of BP by adding an reinforcement term
controlled by the constants $\gamma_0, \gamma_1 \in [0,1]$ to the
marginal $L$-value calculated from the variable nodes (VND) of the
graph-based code. The main modification of our RBP algorithm is
the a priori information from the source. Taking quantizer $C_1$ as
an example. We let $u_i$ be the $i$th element of $C_1$'s input
(uniform), and $c_{1,i}$ be the $i$th coded symbol of codeword
$\mathbf{c_1}$. To reflect the modulo $A_{\epsilon}$ operation
before $C_1$ in Fig.~\ref{SQ_structure}, we use the following
conditional probability density function (PDF) to calculate the a
priori information
\vspace{-0.1cm}
\begin{equation} \label{eq_priori}
f_{u_i|c_{1,\;i}}=\sum_{b\in\mathbb{Z}}f_{G,\sigma_{n,\epsilon}^2}(u_i-c_{1,i}+A_{\epsilon}b),
\end{equation}
\vspace{-0.1cm}
where $f_{G,\sigma_{n,\epsilon}^2}(g)$ is the PDF of $N(0,\sigma_{n,\epsilon}^2)$, and $\mathbb{Z}$ is the integer set.
The rest of the algorithm is the same as that in
\cite{Braunstein_arXiv_11} and is omitted here. After obtaining
$D_{2,\epsilon}$ in Step 4, we can follow Step 5, which is
described in Sec. \ref{Sec_problems}, to get the practical modulo size
$A_p$ in Fig.~\ref{graph} and complete our design.

Finally, note that the complexity of the BP channel decoding algorithm for our WZC
decoder is linear \cite{mackay1999good} in codelength (i.e.
$O(n)$), also is the RBP algorithm used in our WZC encoder. Adopting the RBP algorithm instead of the $O(n^2)$
hard-decimation-based one in \cite{Filler_LDGM}\cite{wang2007approaching} significantly reduces
our computation complexity.

\begin{table}
\caption{Design Flow of Residual WZC} \label{tab:AlgBasic}
\centering \vspace{-2.5mm}
\renewcommand{\arraystretch}{1.25}
\begin{tabular}{ll}
\hline \hline  1: & \textit{Determine the target (ideal) MSE
distortion $D$:}
\\ & Given $P_V$ and WZC rate $R_2$, determine $D$ from \eqref{R_WZC}\\
2: & \textit{Determine code rate $R_1$ of the 1st-stage
quantizer:} \\

&Given $\epsilon$, find $A=A_\epsilon$ such that
$\epsilon_1<\epsilon$ from \eqref{eq_pra_A1}. \\
& Compute $R_1$ from \eqref{Rq_enc} with $A=A_\epsilon$.\\
3: & \textit{Find the maximum tolerable noise variance
$\sigma^2_{n,\epsilon}$ of the channel}\\  & \textit{decoder with
$A=A_\epsilon$:}  \\ & Design SSC LDPC with rate $R_1$. Calculate
$\sigma^2_{n,\epsilon}$ with BP.
\\
4: & \textit{Find the distortion $D_{2,\epsilon}$ of the 2nd-stage
$C_2$ with $A=A_\epsilon$} \\
& Design LDGM with rate $R_2$. \\
& Calculate $D_{2,\epsilon}$ with modified RBP applied to LDPC and LDGM. \\

5: & \textit{Determine the practical modulo size $A_p$:} \\

& Calculate $A_p$ using the RHS of \eqref{Ap}. If the LDPC decoder
fails, \\ & increase $A_p$ until it succeeds. Test the final MSE
with the final $A_p$.
 \\
\hline \hline
\end{tabular}
\vspace{-0.55cm}
\end{table}
\vspace{-0.01cm}
\section{Design example and Discussions}
In our design example, different from the Gaussian side information
in \cite{Xiong_TCOM_09}\cite{DISCUS_IT_03}, we let each element of
the side information $\by$ in \eqref{system_model} be uniformly
distributed in $[-A/2, A/2]$. Due to the dither, the
distribution of $\by$ will not affect our performance.
The details of our design example is given in Sec.
\ref{sec_simulation}, with the distortion performance given in the
end of this subsection. Finally, Sec. \ref{sec_discuss} provides
more discussions on our work.
\vspace{-0.25cm}

\subsection{Design example with detailed code parameters
selection} \label{sec_simulation} Following our design flow in
Table \ref{tab:AlgBasic}, we first set $P_V$ in \eqref{eq_WZ_rate}
as $0.28$ and the WZC rate $R_2=0.953$, and then the ideal
distortion $D$ is 0.0747. For Step 2, given $\epsilon$=0.005, we
find that choosing $A_\epsilon$ as $3$ is sufficient.
$R_1$ is $0.68$ bpcu. For Step 3, we use 2-PAM LDPC to
implement $C_1$ with constellation points
$\pm\frac{A_\epsilon}{4}$. To achieve the WZC bound, we choose
codeword length $n=10^5$ symbols per source block. The degree
profile is :  check nodes (CND): $100\%$ of degree 12; VND:
$35.36\%$ of degree 2, $44.74\%$ of degree 3, and $19.89\%$ of
degree 9. By applying the BP channel decoding algorithm, we obtain
$\sigma^2_{n,\epsilon}$ as 0.185.

For Step 4, we adopt 4-ary LDGM where every two bits of the LDGM CND
are Gray-mapped to a 4-ary symbol. The degree profile can be found
in \cite{wang2007approaching}. By running the RBP quantization
algorithm with $\gamma_0=1$, $\gamma_1=0.99980$ at the LDPC and
LDGM, we obtain $D_{2,\epsilon}$ as 0.0577. The Monte Carlo
simulation tests 2000 blocks of uniformly-distributed source
samples and 3000 iterations are run for each source block.
The summation over integer set $\mathbb{Z}$ in \eqref{eq_priori}
is obtained by limiting $|b|\leq 3, b \in \mathbb{Z}$. As in
\cite{Braunstein_arXiv_11}, in some few cases the RBP algorithm
diverges, i.e. not all the constraints of CND are satisfied. For these few cases, as in
\cite{Braunstein_arXiv_11}, restarting the RBP by adding 0.00001 to $\gamma_1$
will solve the problem.
Finally, from Step 5, we calculate the lower-bound of $A_p$ in
\eqref{Ap} as 3.261. The bit error rate  of LDPC is smaller
than $10^{-4}$ when the modulo size $A_p$ is 3.29.

Finally, we run the Monte Carlo simulation to test 2000 source blocks.
The distortion loss compared with the ideal $D$ is 0.995 dB at the WZC
rate 0.9531 b/s. The best simulation result known with $P_V=0.28$ \cite{Xiong_TCOM_09}, which is only for the
Gaussian side information, has 1.07 dB loss at the rate 1.07 b/s and block length $n=10^5$. The results in \cite{Xiong_TCOM_09} are obtained using
8192-state trellis coded quantization (TCQ) and 3-stage SWC (three LDPCs in the SWC encoder). Our work is as competitive as \cite{Xiong_TCOM_09}. Besides,
our SSC LDPC has only 0.48 dB SNR loss compared to the capacity and
0.43 dB shaping loss compared to the rate-distortion bound.
More design examples can be found in \cite{WZC_journal}.
\vspace{-0.2cm}

\subsection{Detailed comparison} \label{sec_discuss}
Although the well-known practical WZC design \cite{Xiong_TCOM_09}
also performs a two-stage quantization/cpmpression process, the
second stage quantizer/compressor in \cite{Xiong_TCOM_09} is
\textit{lossless} with input being the \textit{quantization
output} of the first stage. In our residual WZC, the second stage
quantizer is \textit{lossy} with input being the
\textit{quantization error} of the first stage, and thus the
decoder structure is also different from that in
\cite{Xiong_TCOM_09}. Compared with the WZC in
\cite{Xiong_TCOM_09}, the main advantage of our coding scheme is
the flexibility. Firstly, as shown in \cite{Lin_ITW_09}, our
residual WZC can approach the Wyner-Ziv bound with arbitrary
side information in all rate regimes, while the WZC in
\cite{Xiong_TCOM_09}\cite{Xiong_IT_06} can only guarantee the
optimality in high rate regimes. In \cite{Xiong_TCOM_09}, the
lower the WZC rate the more severe the rate loss is. Secondly,
our construction offers more flexibility in the practical code rate
selection. In \cite{Xiong_TCOM_09}, the WZC is implemented by a
rate $R_{N}$ TCQ and a multilevel SWC (formed by $R_{N}$ LDPCs).
The rate of TCQ $R_{N}$ is limited to be an integer, and for each
$R_{N}=1,2,\ldots$, only one possible WZC rate can be implemented.
However, almost all the rational WZC rates can be
implemented by our construction.

Our construction is also different from and more flexible than
the graph-based construction in \cite{SCS_IT_09} for which the
encoder codebook is nested.
Moreover, the WZC in \cite{SCS_IT_09} has only been proven to be
optimal for the binary symmetric source with Hamming distortion measure.
Whether it can be extended to the quadratic-Gaussian case as
considered in this paper or not is still unknown. Finally, our
simulation results show that LDPC itself suffices to be a good
SSC code, thus the compound LDGM/LDPC
construction proposed in \cite{SCS_IT_09} may not be necessary for our first-stage
quantizer $C_1$.

\section{Conclusion}
In this paper, we considered the quadratic-Gaussian
Wyner-Ziv problem where side information can be arbitrarily
distributed. We implemented the theoretically-claimed residual WZC by LDPC and LDGM.
We identified and solved a problem called the channel decoding beyond
capacity problem when designing our practical decoder.
Moreover, we modified the RBP algorithm to make the LDPC
with the edge degree optimized for the channel coding perform well as a source code.
The simulation results showed that our practical construction approaches the Wyner-Ziv bound,
and has a similar performance compared with previous works. Moreover, our construction can offer more design
flexibility in terms of the distribution of the side information and the
practical code rate selection.
\vspace{-0.2cm}

\bibliographystyle{IEEEtran}
\renewcommand{\baselinestretch}{0.7}
\bibliography{IEEEabrv,reference_jimtm,CodeSN}

\begin{thebibliography}{10}
\providecommand{\url}[1]{#1}
\csname url@rmstyle\endcsname
\providecommand{\newblock}{\relax}
\providecommand{\bibinfo}[2]{#2}
\providecommand\BIBentrySTDinterwordspacing{\spaceskip=0pt\relax}
\providecommand\BIBentryALTinterwordstretchfactor{4}
\providecommand\BIBentryALTinterwordspacing{\spaceskip=\fontdimen2\font plus
\BIBentryALTinterwordstretchfactor\fontdimen3\font minus
  \fontdimen4\font\relax}
\providecommand\BIBforeignlanguage[2]{{%
\expandafter\ifx\csname l@#1\endcsname\relax
\typeout{** WARNING: IEEEtran.bst: No hyphenation pattern has been}%
\typeout{** loaded for the language `#1'. Using the pattern for}%
\typeout{** the default language instead.}%
\else
\language=\csname l@#1\endcsname
\fi
#2}}

\bibitem{Wyner-t2}
A.~D. Wyner, ``The rate-distortion function for source coding with side
  information at the decoder-$\textsc{II}$: General sources,'' \emph{Inf.
  Contr.}, vol.~38, pp. 60--80, Jan. 1978.

\bibitem{Slepian-n}
D.~Slepian and J.~K. Wolf, ``Noiseless coding of correlated information
  sources,'' \emph{{IEEE} Trans. Inform. Theory}, vol.~19, no.~4, pp. 471--480,
  July 1973.

\bibitem{Xiong_SPM_04}
Z.~Xiong, A.~D. Liveris, and S.~Cheng, ``Distributed source coding for sensor
  networks,'' \emph{{IEEE} Signal Processing Mag.}, vol.~21, no.~5, pp. 80--94,
  Sep. 2004.

\bibitem{Zamir-n}
R.~Zamir, S.~Shamai(Shitz), and U.~Erez, ``Nested linear/lattice codes for
  structured multiterminal binning,'' \emph{{IEEE} Trans. Inform. Theory},
  vol.~48, no.~6, pp. 1250--1276, Jun. 2002.

\bibitem{Kramer-c}
G.~Kramer, M.~Gastpar, and P.~Gupta, ``Cooperative strategies and capacity
  theorems for relay networks,'' \emph{{IEEE} Trans. Inform. Theory}, vol.~51,
  no.~9, pp. 3037--3063, Sep. 2005.

\bibitem{SCS_IT_09}
M.~J. Wainwright and E.~Martinian, ``Low-density graph codes that are optimal
  for binning and coding with side information,'' \emph{{IEEE} Trans. Inform.
  Theory}, vol.~55, no.~3, pp. 1061--1079, Mar. 2009.

\bibitem{Korada_IT_10}
S.~B. Korada and R.~L. Urbanke, ``Polar codes are optimal for lossy source
  coding,'' \emph{{IEEE} Trans. Inform. Theory}, vol.~56, no.~4, pp.
  1751--1768, Apr. 2010.

\bibitem{Lin_ITW_09}
S.-J. Lin, S.-C. Lin, K.-S. Chen, and H.-J. Su, ``Coding for noisy
  quadratic-{G}aussian {W}yner-{Z}iv problem: {A} successive quantization
  approach,'' in \emph{IEEE Information Theory Workshop}, 2009.

\bibitem{gersho1992vector}
A.~Gersho and R.~M. Gray, \emph{Vector quantization and signal
  compression}.\hskip 1em plus 0.5em minus 0.4em\relax Kluwer Academic
  Fublishers, 1992.

\bibitem{Xiong_TCOM_09}
Y.~Yang, S.~Cheng, Z.~Xiong, and W.~Zhao, ``Wyner-{Z}iv coding based on {TCQ}
  and {LDPC} codes,'' \emph{{IEEE} Trans. Commun.}, vol.~57, no.~2, pp.
  376--387, Feb. 2009.

\bibitem{Xiong_IT_06}
Z.~Liu, S.~Cheng, A.~D. Liveris, and Z.~Xiong, ``Slepian-{W}olf coded nested
  lattice quantization for {W}yner-{Z}iv coding: {H}igh-rate performance
  analysis and code design,'' \emph{{IEEE} Trans. Inform. Theory}, vol.~52,
  no.~10, pp. 4358--4379, Oct. 2006.

\bibitem{Chandar_thesis_10}
V.~Chandar, ``Sparse graph codes for compression, sensing, and secrecy,'' Ph.D.
  dissertation, Massachusetts Institute of Technology, June 2010.

\bibitem{Braunstein_arXiv_11}
\BIBentryALTinterwordspacing
A.~Braunstein, F.~Kayhan, and R.~Zecchina, ``Efficient data compression from
  statistical physics of codes over finite fields,'' \emph{Phys. Rev. E},
  vol.~84, p. 051111, Nov 2011. [Online]. Available:
  \url{http://link.aps.org/doi/10.1103/PhysRevE.84.051111}
\BIBentrySTDinterwordspacing

\bibitem{Filler_LDGM}
T.~Filler and J.~Fridrich, ``Binary quantization using belief propagation with
  decimation over factor graphs of {LDGM} codes,'' in \emph{45th Annual
  Allerton Conf. on Comm. Control, and Comput., Monticello, IL, USA}, 2007.

\bibitem{Chen-s}
K.-S. Chen, ``Low density constructions for simultaneously good for channel and
  source coding problem with applications,'' Master's thesis, National Taiwan
  university, 2009.

\bibitem{mackay1999good}
D.~J.~C. MacKay, ``Good error-correcting codes based on very sparse matrices,''
  \emph{{IEEE} Trans. Inform. Theory}, vol.~45, no.~2, pp. 399--431, March
  1999.

\bibitem{Brink_TCOM_04}
S.~ten Brink, G.~Kramer, and A.~Ashikhmin, ``Design of low-density parity-check
  codes for modulation and detection,'' \emph{{IEEE} Trans. Commun.}, vol.~52,
  no.~4, pp. 670--678, Apr. 2004.

\bibitem{wang2007approaching}
Q.~Wang and C.~He, ``Approaching 1.53-d{B} shaping gain with {LDGM}
  quantization codes,'' in \emph{IEEE Global Telecommunications Conference},
  2007, pp. 1571--1576.

\bibitem{DISCUS_IT_03}
S.~S. Pradhan and K.~Ramchandran, ``Distributed source coding using syndromes
  ({DISCUS}): {D}esign and construction,'' \emph{{IEEE} Trans. Inform. Theory},
  vol.~49, no.~3, pp. 626--643, Mar. 2003.

\bibitem{WZC_journal}
Y.-P. Wei, S.-C. Lin, S.-J. Lin, and H.-J. Su, ``Residual-quantization based
  code design for source coding with arbitrary decoder side information,''
  \emph{to be submitted to IEEE Trans. Inform. Theory}.

\end{thebibliography}

\end{document}